\documentclass[12pt]{article}
\usepackage[pdftex]{graphicx}
\usepackage{amsmath}
\usepackage{amssymb}

 \usepackage[
    natbib=true,
    style=numeric-comp,
    sorting=none
]{biblatex}

\AtEveryBibitem{\clearfield{title}}    
\AtEveryBibitem{\clearfield{doi}}   
\AtEveryBibitem{\clearfield{pages}}
\renewbibmacro{in:}{}

\usepackage{multirow}

\newcommand {\E}[1]{\times 10^{#1}}	
\newcommand {\e}[1]{\mathrm{~#1}}       
\newcommand{\mc}[1]{\mathcal{#1}}

\def\support{\footnote{Supported by Slovenian research agency.}}

\def\Title#1{\begin{center} {\Large #1 } \end{center}}
\def\Author#1{\begin{center}{ \sc #1} \end{center}}
\def\Address#1{\begin{center}{ \it #1} \end{center}}

\newenvironment{Abstract}{\begin{quotation}  }{\end{quotation}}
\newenvironment{Presented}{\begin{quotation} \begin{center} 
             PRESENTED AT\end{center}\bigskip 
      \begin{center}\begin{large}}{\end{large}\end{center} \end{quotation}}

\def\beq{\begin{equation}}
\def\eeq#1{\label{#1}\end{equation}}
\def\eeqn{\end{equation}}

\def\beqa{\begin{eqnarray}}
\def\eeqa#1{\label{#1}\end{eqnarray}}
\def\eeqan{\end{eqnarray}}

\let\bar=\overbar

\def\Dslash{\not{\hbox{\kern-4pt $D$}}}
\def\dslash{\not{\hbox{\kern-2pt $\del$}}}

\def\msb{{\bar{\ssstyle M \kern -1pt S}}}

\begin{document}
\begin{titlepage}

\vfill
\Title{New Physics Models Facing Lepton Flavor Violating Higgs Decays}
\vfill
\Author{Nejc Ko\v snik\support}
\Address{Department of Physics, University of Ljubljana, Jadranska 19,
  1000 Ljubljana, Slovenia\\ 
and\\
Jo\v zef Stefan Institute, Jamova 39, P.O. Box 3000, 1001 Ljubljana, Slovenia}
\vfill
\begin{Abstract}
We speculate about the possible interpretations of the recently
observed excess in the $h \to \tau \mu$ decay. We derive a robust lower
bound on the Higgs boson coupling strength to a tau and a muon, even
in presence of the most general new physics affecting other Higgs
properties. Then we reevaluate complementary indirect constraints
coming from low energy observables as well as from theoretical
considerations. In particular, the tentative signal should lead to $\tau
\to \mu \gamma$ at rates which could be observed at Belle II. In turn we
show that, barring fine-tuned cancellations, the effect can be
accommodated within models with an extended scalar sector. These
general conclusions are demonstrated in explicit new
physics models. Finally we show how, given the $h \to \tau \mu$ signal, the
current and future searches for $\mu \to e \gamma$ and $\mu \to e$ nuclear
conversions unambiguously constrain the allowed rates for $h \to \tau e$.
\end{Abstract}
\vfill
\begin{Presented}
The 7th International Workshop on Charm Physics (CHARM 2015)\\
Detroit, MI, 18-22 May, 2015
\end{Presented}
\vfill
\end{titlepage}
\def\thefootnote{\fnsymbol{footnote}}
\setcounter{footnote}{0}
%

\section{Introduction}


The discovery of the Higgs boson~\cite{Aad:2012tfa,
  Chatrchyan:2012ufa} offers numerous observables where the validity
of the Standard Model~(SM) can be tested. The CMS collaboration has
recently reported a slight excess with a significance of $2.4\,\sigma$
in the search for LFV decay
$h\to \tau \mu$~\cite{Khachatryan:2015kon}. The best fit for the
branching ratio of the Higgs boson to $\tau \mu$ under assumption of
SM Higgs production is found to be
\begin{equation}
\label{eq:Br}
\mathcal{B}(h\to \tau \mu)=\left(0.84^{+0.39}_{-0.37}\right) \% \,.
\end{equation}
This recent hint has received great attention in the
literature~(see~\cite{Dorsner:2015mja} and references therein).
An ATLAS study of $h \to \tau \mu$~\cite{Aad:2015gha} in the hadronic $\tau$ channel
observes no excess and is consistent with the CMS result~\eqref{eq:Br}.

In light of the tentative signal it is instructive
to revisit compatibility of such large $\mc{B}(h \to \tau \mu)$ with
other low-energy lepton flavor violation~(LFV) probes in the context
of popular New physics~(NP) models.

\section{Higgs effective theory approach}
\label{sec:EFT}
The mass terms and Higgs boson couplings of charged
leptons after electroweak symmetry breaking (EWSB) can be
parametrized in the general case as
\begin{equation}
\mathcal L^{\rm eff.}_{Y_\ell} = - m_i \delta_{ij} \bar \ell_L^i \ell_R^j - y_{ij}\left( \bar \ell_L^i \ell_R^j\right) h + \ldots+ \rm h.c. \,,
\label{eq:LY}
\end{equation}
where the ellipsis denotes non-renormalizable terns involving more
than one Higgs boson. While in the SM the Yukawas $y \sim m$ are diagonal in the mass
basis, NP could misalign the two matrices and via $y_{\tau \mu}$ and/or $y_{\mu \tau}$ induce $h\to \tau \mu$ decays with a  branching ratio of
$\mathcal{B}(h\to \tau \mu) = \frac{m_h}{8\pi \Gamma_{h}} \left(|y_{\tau \mu}|^2+|y_{\mu \tau}|^2\right)$.
If one assumes that the total Higgs boson decay width ($\Gamma_{h}$)
is SM-like and enlarged only by the contribution from $h\to \tau\mu$
then the measurement in Eq.~\eqref{eq:Br} can be interpreted as a
two-sided bound,
$0.0019 (0.0008) < \sqrt{|y_{\tau \mu}|^2+|y_{\mu \tau}|^2} < 0.0032
(0.0036)$  at 68\% (95\%) C.L..
In general, the experimentally measured $h \to \tau\mu$ branching fraction
depends also on other Higgs couplings contributing both to its total decay
width ($\Gamma_{h}$) as well as its production cross-section
($\sigma_h$). In particular, a given signal can be reproduced  for
larger (smaller) values of $|y_{\tau\mu}|$ and $|y_{\mu\tau}|$ by
enhancing (suppressing) $\Gamma_{h}$ and/or suppressing (enhancing)
$\sigma_h$. 
Individual effects of $\Gamma_h$ and $\sigma_h$ can
be disentangled by performing a global fit to all Higgs production and
decay event yields at the LHC~\cite{Dorsner:2015mja}. 
Numerically, we find in this case
\begin{equation}
0.0017 (0.0007) < \sqrt{|y_{\tau \mu}|^2+|y_{\mu \tau}|^2} < 0.0036 (0.0047) ~ {\rm at}~ 68\%~ (95\%)~{\rm C.L.}\,.
\label{eq:htaummu1b}
\end{equation}

In the following we assume the SM contains all the relevant degrees of
freedom at energies $\mathcal O({\rm few~} 100)$\,GeV, whereas
for the additional heavy degrees of freedom we assume they been
integrated out. The natural ranges for the effective Higgs couplings follow from the hierarchy between the
muon and tau lepton masses~\cite{Cheng:1987rs, Branco:2011iw},
\begin{equation}
\label{eq:lfv-nat}
\sqrt{|y_{\tau\mu} y_{\mu\tau}|} \lesssim \frac{\sqrt{m_\mu m_\tau}}{v} = 0.0018\,.
\end{equation} 
It is interesting to note that almost the whole parameter space in
allowed by $h \to \tau \mu$ is also compatible with the above
naturalness criterium~\cite{Dorsner:2015mja}.

\begin{figure}[!h]
\centering \begin{tabular}{cc}
\includegraphics[width=0.44\textwidth]{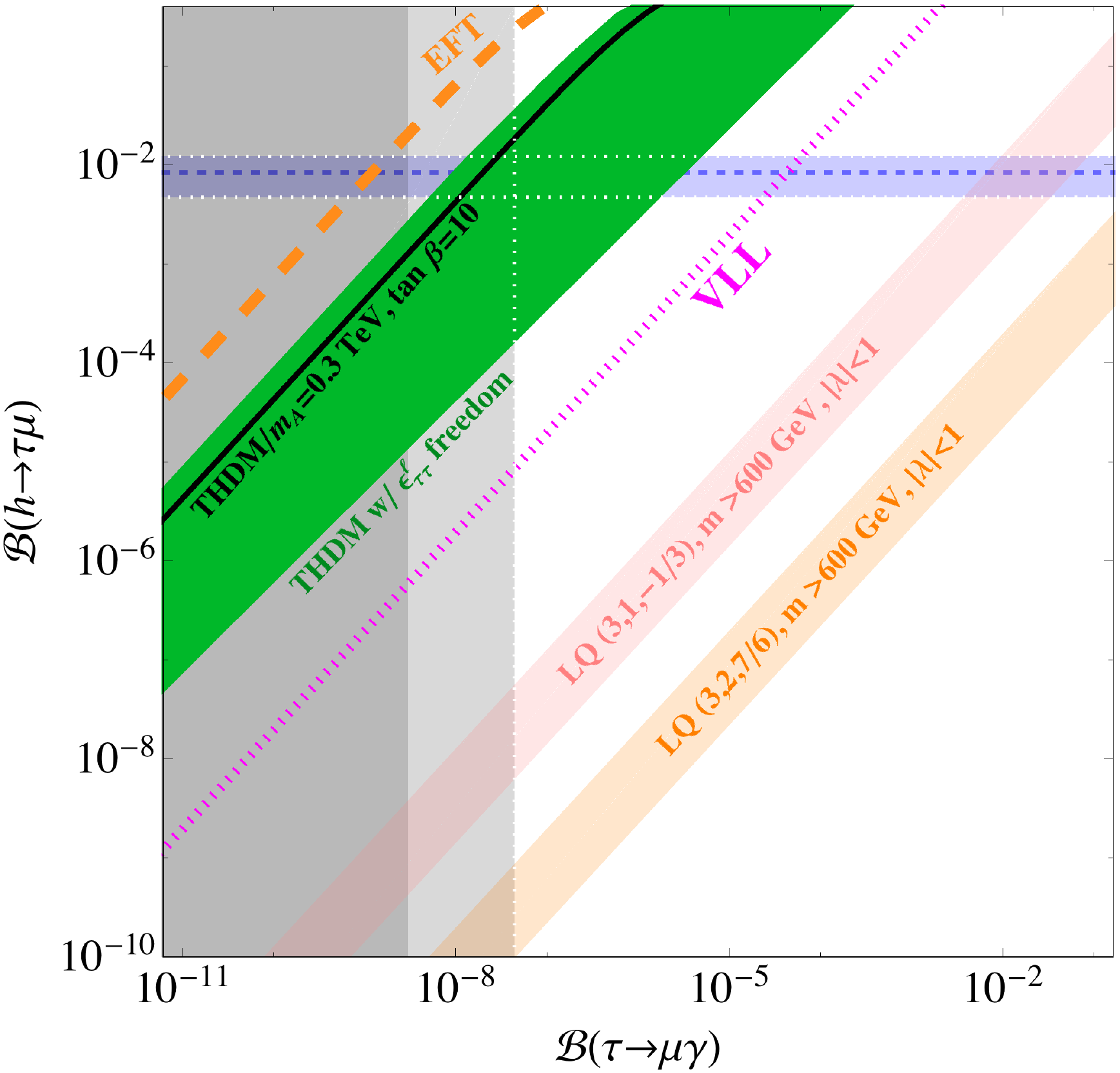}
             &\includegraphics[width=0.43\textwidth]{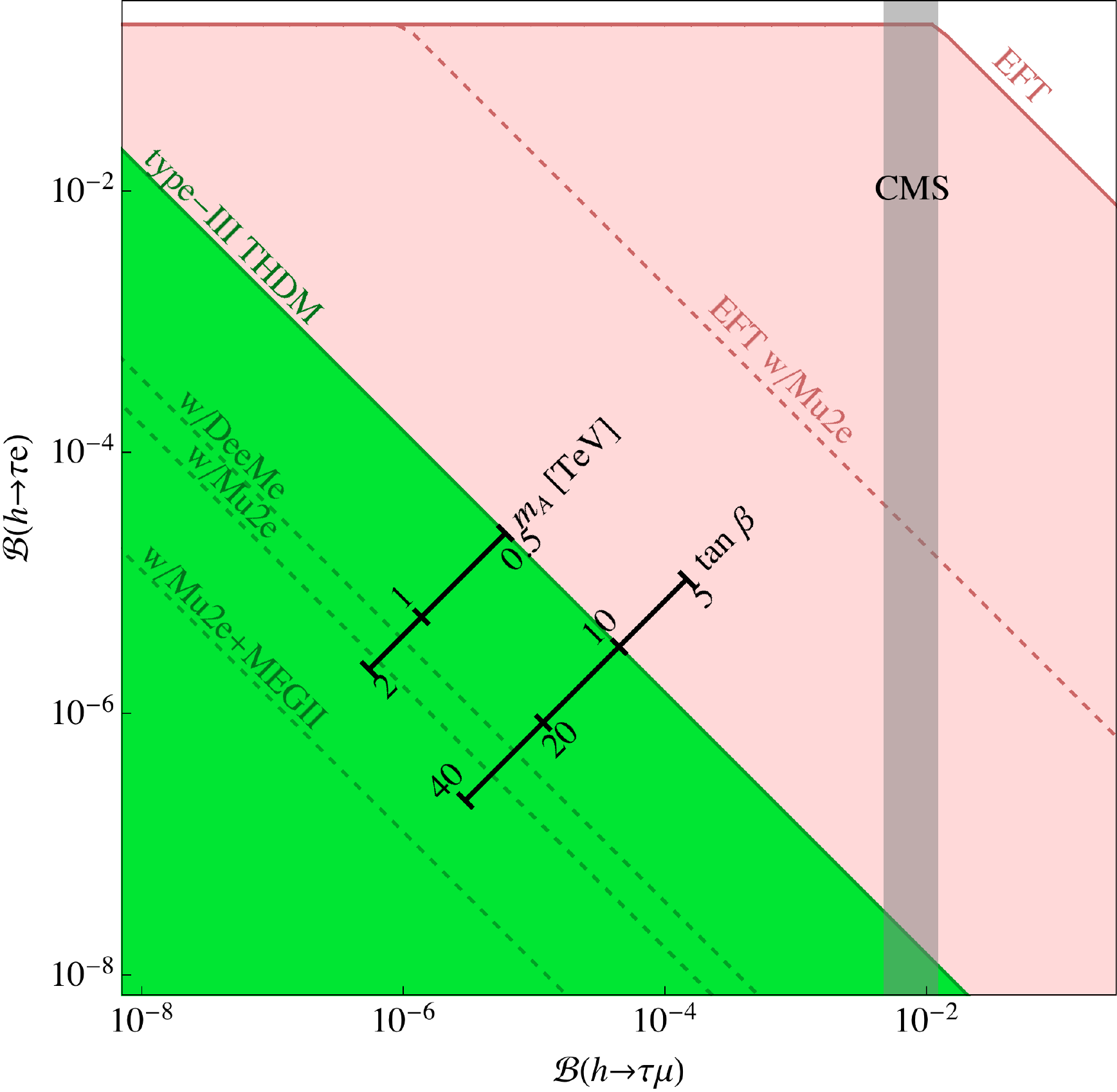}
\end{tabular}
\caption{Left-hand side panel: Correlation between $\mathcal B(h\to \tau\mu)$ and $\mathcal
  B(\tau \to \mu\gamma)$ in various NP scenarios. The present
  experimental result for $\mathcal B(h\to \tau\mu)$ is shown in
  horizontal blue band~\cite{Khachatryan:2015kon}. Current and future
  projections for $\mathcal B(\tau \to \mu\gamma)$ experimental
  sensitivity are represented with vertical light~\cite{Aubert:2009ag}
  and dark~\cite{Aushev:2010bq} gray bands, respectively. Superimposed
  are the predictions within the EFT approach (diagonal dashed orange
  line), in the type-III THDM (green and black bands), and in models with
  scalar leptoquarks (diagonal red and orange shaded band). Right-hand
  side panel: Allowed region in the $\mc{B}(h \to \tau e)$--$\mc{B}(h \to \tau \mu)$ plane when experimental upper bounds on $\mu \to e \gamma$ and $\mu - e$ conversion rates are taken into account. Pink region is permitted in the effective theory setting while the dashed line indicates how much the region will shrink if Mu2e and MEG II experiments see no signal events. Green region is allowed within type-III THDM
model with $m_A = 0.5\e{TeV}$ and $\tan \beta = 10$. Rulers indicate how much the region shrinks with increasing $\tan \beta$ or $m_A$, while dashed lines correspond to improved experimental upper bounds on $\mu \to e \gamma$ and $\mu-e$ as described in the text.}
\label{fig:master}
\end{figure}

Phenomenologically the most relevant low-energy constraints on this
scenario are the
one- and two-loop contributions to the operators
$\bar L H (\sigma \cdot B) E$ and
$\bar L \tau_a H (\sigma \cdot W^a) E$, where
$\sigma_{\mu\nu} = i[\gamma_\mu,\gamma_\nu]$/2, $B_{\mu\nu}$ and
$W^a_{\mu\nu}$ are the hypercharge and weak isospin field strengths,
respectively, and $\tau_a$ are the Pauli matrices. These operators can
mediate the strongly constrained radiative LFV decays. The most stringent constraint comes from the
$\tau \to \mu\gamma$ decay~\cite{Harnik:2012pb}, mediated by the
effective Lagrangian
\begin{equation}
\mathcal L_{\rm eff.} = c_L \mathcal Q_{L\gamma} + c_R \mathcal Q_{R\gamma} + \rm h.c.\,,
\label{eq:Leff}
\end{equation}
where $\mathcal Q_{L,R\gamma} = (e/8\pi^2) m_\tau (\bar \mu
\sigma^{\alpha\beta} P_{L,R} \tau)F_{\alpha\beta}$, $P_{L,R} = (1\mp
\gamma_5)/2$ and $F_{\alpha\beta}$ is the electromagnetic field
strength tensor.
The resulting EFT correlation between $\mathcal B(h\to \tau \mu)$ in
and $\mathcal B(\tau \to \mu\gamma)$ 
is shown in left-hand panel in Fig.~\ref{fig:master} (diagonal dashed orange line), assuming SM values of all Higgs boson couplings except $y_{\tau\mu}$ and $y_{\mu\tau}$.
In the same plot, the CMS preferred range of $\mathcal B(h\to\tau\mu)$
in Eq.~\eqref{eq:Br} is displayed by the horizontal blue band, while
the current ($\mathcal B(\tau\to\mu\gamma) < 4.4 \times
10^{-8}$~@~90\%~C.L.)~\cite{Aubert:2009ag} and projected future
($\mathcal B(\tau\to\mu\gamma) < 3 \times
10^{-9}$~@~90\%~C.L.)~\cite{Aushev:2010bq} indirect constraints are
shaded in light and dark gray vertical bands, respectively. We observe
that within the EFT approach, the CMS signal is well compatible with
the non-observation of $\tau\to\mu\gamma$ at the $B$ factories and will
marginally remain so even at Belle~II. 
\section{Type-III Two Higgs Doublet Model}
\label{sec:THDM}
Extensions of the SM with an additional $SU(2)_L$ doublet scalar are an effective description of several
NP scenarios (e.g. supersymmetric extensions or models addressing the
strong CP problem via a Peccei-Quinn symmetry), for a recent review c.f.~\cite{Branco:2011iw}. The most extensively studied version of the
Two Higgs Doublet Model~(THDM) is the type-II model in which one of the doublets couples to
up-type quarks, while the other one couples to down-type quarks and charged
leptons, avoiding in this way the tree level flavor changing neutral
currents (FCNCs). All LFV in this case comes at 1-loop and is aligned
with the SM flavor structure thus rendering LFV processes to a
negligible level, compared to the one of the SM~\cite{Dorsner:2015mja}.
In type-III THDM Yukawa couplings are generic and allow for large LFV.
In the case with a MSSM-like Higgs potential the following relations hold~\cite{Crivellin:2013wna}
\begin{equation}
\begin{split}
\tan\beta &= \frac{v_{u}}{v_{d}},\quad \tan 2\alpha= \tan 2\beta\frac{m_{A}^{2}+m_{Z}^{2}}{m_{A}^{2}-m_{Z}^{2}}\,,
\end{split}
\end{equation}
while for the masses $m_{H^{\pm}}^{2} = m_{A}^{2}+m_{W}^{2}$ and $m_{H}^{2} = m_{A}^{2}+m_{Z}^{2}-m_{h}^{2}$.
Here $\beta$ is the angle that diagonalizes the mass matrices
of charged scalars and pseudoscalars while $\alpha$ is an analogous angle for the
neutral scalars. 
The relevant part for the discussion of LFV are the Yukawa couplings in the charged lepton mass eigenstate basis~\cite{Crivellin:2013wna}
\begin{equation}
\mathcal{L} =\frac{y_{fi}^{H_k}}{\sqrt{2}}H_k
\bar{\ell}_{L,f}\ell_{R,i}+ \frac{y_{fi}^{H^+}}{\sqrt{2}} H^+
\bar{\nu}_{L,f}\ell_{R,i} + \textrm{h.c.}\,,
\end{equation}
where the LFV Yukawa couplings can be written as
\begin{equation}
y_{fi}^{H_{k}}=x_{d}^{k}\frac{m_{\ell_{i}}}{v_{d}}\delta_{fi}-\epsilon_{fi}^{\ell}\left(x_{d}^{k}\,\tan\beta-x_{u}^{k*}\right)\,.
\end{equation}
The off-diagonal parameters $\epsilon^\ell_{fi}$ drive the LFV
phenomena, while the coefficients $x_q^k$ for $H_k=(H,h,A)$ can be
found in~\cite{Dorsner:2015mja}.
Using the above relations, we find the tree-level result for the
$h \to \tau \mu$ branching fraction,
\begin{equation}
\label{eq:THDM-Brh}
  \mc{B}(h\to \tau \mu) = \frac{m_h}{16 \pi \Gamma_h}\,(\sin \alpha \tan\beta + \cos \alpha)^2\,\left(|\epsilon_{\mu\tau}^\ell|^2 + |\epsilon_{\tau\mu}^\ell|^2\right)\,.
\end{equation} 
We have found that even for light pseudoscalar masses  ($m_A$),
modifications of $hWW$ and $hZZ$ relative to their SM values is negligible~\cite{Dorsner:2015mja}.


The decay width of $\tau \to \mu \gamma$ is driven by
a one- and two-loop Barr-Zee diagrams with virtual charged or neutral
Higgses. At one-loop the amplitude is suppressed by two small Yukawa
couplings, while the two-loop result is proportional to $y_{tt}$ and a
single LFV Yukawa. Indeed one finds that Barr-Zee contributions
dominate the rate of $\tau \to \mu \gamma$~\cite{Dorsner:2015mja}.
We sample the parameter space of $\epsilon^\ell_{\tau \mu}$,
$\epsilon^\ell_{\mu \tau}$, $\epsilon^\ell_{\tau\tau}$ for a chosen values of $\tan \beta$ and $m_A$.
The ranges allowed for $\epsilon^\ell_{\tau \mu,\mu\tau}$ are required to fulfill the naturalness criterium of Eq.~\eqref{eq:lfv-nat}
and stay within the perturbative regime.

The 1-loop amplitude of the $\tau \to \mu \gamma$ process further depends on the diagonal $y_{\tau\tau}^h$ Yukawa coupling  which is experimentally
constrained by the searches for $h \to \tau\tau$ decays by the
ATLAS~\cite{Aad:2015vsa} and CMS~\cite{Chatrchyan:2014nva}
experiments. The naively averaged signal strength of two $\tau\tau$
signal strengths results in $\mu^{\tau\tau} = 1.02^{+0.21}_{-0.20}$
which directly constrains $y_{\tau\tau}^h$. A scenario with SM-like
$y^h_{\tau\tau}$ coupling corresponding to $\mu^{\tau\tau} = 1$ for
fixed $\tan \beta = 10$ and $m_A = 0.3\e{TeV}$  is presented in
left-hand side panel in Fig.~\ref{fig:master} by a black narrow stripe. This scenario easily
passes both experimental constraints. 
On the other hand, for masses $m_A$ significantly larger than
$500\e{GeV}$ it is not possible to reconcile both predictions with the
corresponding experimental values. Allowing $\epsilon_{\tau\tau}^\ell$
to departure from zero within the allowed range we obtain
correlation represented by the green band in left-hand side panel in Fig.~\ref{fig:master}. In
particular, this additional freedom in $\tau \to \mu \gamma$ breaks
the strict correlation with the $h \to \tau \mu$ rate. 

\section{Scalar Leptoquarks}
\label{sec:LQ}
A scalar leptoquark state (LQ) can induce $h \to \tau \mu$ decay via
quark-LQ penguin diagrams. 
Inspection of the helicity structure of the relevant diagrams reveals
that both chiralities of leptons and top quark have to couple to the
LQ state, in order keep the leptoquark couplings perturbative~\cite{Dorsner:2015mja}.
LQ state can couple to the Higgs also via the ``Higgs portal" operator,
$-\lambda H^{\dagger}H\Delta^{\dagger}\Delta$ that
comes with an additional parameter, $\lambda$.

\subsection{The $\Delta_1=(\bf{3},\bf{1},-1/3)$ case}
The Yukawa couplings of $\Delta_1$ are given by the following Lagrangian
\begin{equation}
\mathcal{L}_\mathrm{\Delta_1}=y^L_{ij}\bar{Q}_{}^{i,a}\Delta_1\epsilon^{ab}L_{}^{C\,j,b}+y^R_{ij}\bar{U}_{}^{i}\Delta_1 E_{}^{C\,j}+\textrm{h.c.}\,,
\end{equation}
where $Q^i=(u^i_L,d^i_L)^T$ and $U^i=u^i_R$ are the quark weak
doublets and up-quark singlets, respectively.  We explicitly show
flavor indices $i,j=1,2,3$, and $SU(2)$ indices $a,b=1,2$, with
$\epsilon_{12} = 1$. 
The  $h \to \tau \mu$ decay width in the presence of $\Delta_1$ scalar
is then,
\begin{equation}
\label{eq:LQ1-BR}
\Gamma(h\to\tau\mu)=\frac{9 m_{h}m_{t}^{2}}{2^{13}\pi^{5}v^{2}}\left(\left|y_{t\mu}^{L}y_{t\tau}^{R}\right|^{2}+\left|y_{t\tau}^{L}y_{t\mu}^{R}\right|^{2}\right)\left|g_1(\lambda,m_{\Delta_1})\right|^{2}\,.
\end{equation}
Here the relevant loop function further depends on the portal coupling
$\lambda$ and is given in~\cite{Dorsner:2015mja}.
The state $\Delta_1$ also contributes to the $\tau \to \mu \gamma$ via
same $y$ couplings as the ones present in $h \to \tau \mu$:
\begin{equation}
 \mathcal{B}(\tau \to \mu \gamma) = \frac{\alpha m_\tau^3}{2^{12}
   \pi^4 \Gamma_\tau}\,\frac{m_t^2}{m_{\Delta_1}^4} \, h_1(x_t)^2 \,
 \left(\left|y_{t\mu}^{L}y_{t\tau}^{R}\right|^{2}+\left|y_{t\tau}^{L}y_{t\mu}^{R}\right|^{2}\right)\,,
\end{equation}
where $x_t=m_t^2/m_{\Delta_1}^2$ and expression for the $h_1$ function
can be found in Ref.~\cite{Dorsner:2015mja}.

The impact of a non-zero Higgs portal coupling
$\lambda$ in the scenario with $\Delta_1$ has also been studied. As an example, for $m_{\Delta_1}=650$\,GeV the
loop function dependence on the portal coupling is
$g_1 =-(0.26+0.12 \lambda)$.
Thus, a positive large $\lambda$ could relax the
leptoquark Yukawa couplings and yield sizable $h\to \tau \mu$ rates
without violating the $\tau \to \mu \gamma$ constraint. However, the
Higgs portal coupling also induces corrections to the
$h\to\gamma\gamma$ decay and to gluon-gluon fusion (ggF) induced Higgs
production with the leptoquark running in the triangular loop.  Taking
those modifications of the Higgs couplings into account we have found
that one can compensate small leptoquark Yukawas by scaling up $\lambda$,
which is itself bounded from above only by perturbativity
requirements~\cite{Dorsner:2015mja}.

The correlation between the $h\to \tau \mu$ and $\tau \to \mu \gamma$
branching ratios in  presence of the $(3,1,-1/3)$ leptoquark state for
$|\lambda| < 1$ and $m_{\Delta_1} > 600\e{GeV}$ is depicted in
left-hand side panel in Fig.~\ref{fig:master} with a pink stripe,
demonstrating that $\tau \to \mu \gamma$ basically excludes this LQ
state as an explanation of $h \to \tau \mu$ signal.

\subsection{The $\Delta_2=(\bf{3},\bf{2},7/6)$ case}
The Yukawa couplings of the $\Delta_2$ leptoquark to  SM fermions are
\begin{equation}
\label{eq:main_2}
 \mathcal{L}_\mathrm{\Delta_2} = y^L_{ij}\bar{E}_{}^{i} \Delta_{2}^{a\,*}Q_{}^{j,a} -y^R_{ij}\bar{U}_{}^{i} \Delta_{2}^{a}\epsilon^{ab}L_{}^{j,b}+\textrm{h.c.},
\end{equation}
The $h \to \tau \mu$ decay rate in this case is
\begin{equation}
  \Gamma(h \to \tau \mu) = \frac{9 m_h m_t^2}{2^{13} \pi^5 v^2} |g_{1}(\lambda, m_{\Delta_2})|^2 \left(
|y_{\mu t}^L y_{t\tau}^R|^2 +
|y_{\tau t}^L y_{t\mu}^R|^2
 \right)\,.
\end{equation}
Allowed values for the Higgs portal coupling $\lambda$ can be inferred
from a global fit to the Higgs data as has been done for the portal
coupling of the $(3,1,-1/3)$ state.

The decay width of $\tau \to \mu \gamma$ are proportional to the couplings responsible for $h \to \tau \mu$:
\begin{equation}
  \label{eq:LQ76-tau-mug}
  \mathcal{B} (\tau \to \mu \gamma) = \frac{\alpha m_\tau^3}{2^{12} \pi^4 \Gamma_\tau} \frac{m_t^2}{m_\Delta^4}\,h_2(x_t)^2\, \left(|y_{t \tau}^R y_{\mu t}^L|^2 + |y_{t \mu}^R y_{\tau t}^L|^2\right)\,.
\end{equation}


In this leptoquark scenario the bound on $\mc{B}(\tau \to \mu
\gamma)$ excludes sizable $\mc{B}(h \to \tau \mu)$ due to the strict
correlation between the two observables. See the orange stripe in
left-hand side panel in Fig.~\ref{fig:master}, where the portal coupling is restricted to $|\lambda| < 1$.

\section{$h \to \tau \mu$ vs.\ $h \to \tau e$}
\label{sec:hlfv-corr-eft}
A positive indication for $h \to \tau \mu$ decay can be combined with
stringent experimental limits on $\mu-e$ LFV processes to constrain
$\tau-e$ processes. In particular, in models with tree-level
$h \to \tau \mu$ (EFT, THDM III) the product of
the $\mc{B}(h \to \tau e)$ and $\mc{B}(h \to \tau \mu)$ is bounded
from above by the rates of $\mu \to e \gamma$ and $\mu-e$ conversion
on nuclei. This is due to the fact that tree level Higgs decays to
$\tau \mu$ ($\tau e$) depend on $y_{\mu \tau, \tau \mu}$
($y_{e \tau, \tau e}$) while the same sets of couplings contribute at
the loop level to $\mu \to e \gamma$ and $\mu-e$ conversion via
diagrams with a virtual $\tau$.  In the effective theory framework,
the contributions to the $\mu \to e \gamma$ process stemming from a
virtual $\tau$ are $m_\tau$ enhanced with respect to diagrams with
intermediate $\mu$ or $e$ states~\cite{Harnik:2012pb} leading to
\begin{equation}
  \label{eq:muegamma-cL}
c_L^{\tau} \simeq \frac{m_\tau}{m_\mu}\, \frac{-3+4 x_\tau - x_\tau^2 - 2 \log x_\tau}{8m_h^2 (1-x_\tau)^3}\,y_{\mu \tau}^* y_{\tau e}^*\,,\qquad x_\tau = \frac{m_\tau^2}{m_h^2}\,,
\end{equation}
where we have neglected the effects of the light lepton masses. The coefficient $c_R^\tau$ is obtained from Eq.~\eqref{eq:muegamma-cL} by replacing $y_{ij} \to y_{ji}^*$. The $\mu \to e \gamma$ branching fraction is thus sensitive to a
distinct combination of the LFV Yukawas:
\begin{equation}
  \label{eq:5}
  \mc{B}(\mu \to e \gamma) \simeq \mc{B}_0^{\mu \to e \gamma} \left(|y_{\mu \tau} y_{\tau e}|^2 + |y_{\tau \mu} y_{e \tau}|^2\right)\,,\qquad \mc{B}_0^{\mu \to e \gamma} = 185\,.
\end{equation}
On the other hand, $\mu-e$ conversion on nuclei is most sensitive to
vector current effective operators $(\bar e \gamma_\nu
P_{L,R}\mu)\,(\bar q \gamma^\nu q)$. 
The branching fraction $\mathcal B(\mu\to e)_{\rm Au} \equiv
{\Gamma(\mu\to e)_\textrm{Au}}/{\Gamma_\textrm{capture Au}}$ can be
put in the form
\begin{equation}
  \label{eq:mueconvEFT}  
\begin{split}
\mathcal B(\mu\to e)_{\rm Au}  
&= \mc{B}_0^{\mu e}\,\left(|y_{e\tau} y_{\mu \tau}|^2 + |y_{\tau e} y_{\tau \mu}|^2\right)\,,\qquad \mc{B}_0^{\mu e} = 4.67\E{-4}\,,
\end{split}
\end{equation}
where the relevant numerics have been taken from
Ref.~\cite{Kitano:2002mt} (also c.f.~\cite{Dorsner:2015mja}).  The complementary information on the LFV couplings extracted from $\mu
\to e \gamma$ and $\mu-e$ conversion allows for correlation between the Higgs LFV  $h \to \tau \mu$ and $h \to \tau e$ decays:
\begin{equation}
\begin{split}
  \label{eq:taue-taumu}
  \mc{B}(h \to \tau \mu) \mc{B}(h \to \tau e) 
&= 8\E{-10}\,\left[\frac{\mc{B}(\mu \to e \gamma)}{10^{-13}}\right] + 3\E{-4}\, \left[\frac{\mathcal B(\mu\to e)_{\rm Au}}{10^{-13}}\right]\,.
\end{split}
\end{equation}
The best experimental limit on $\mathcal B(\mu\to e)_{\rm Au} < 7\E{-13}$~(at 90\% C.L.) was achieved by the SINDRUM II Collaboration~\cite{Bertl:2006up} while the best upper bound on $\mc{B}(\mu \to e \gamma) < 5.7\E{-13}$~(at 90\% C.L.) was recently determined by the MEG Collaboration~\cite{Adam:2013mnn}. Note that with the current experimental data the sum on the right-hand side of Eq.~\eqref{eq:taue-taumu} is completely saturated by the $\mu-e$ conversion contribution. Combining the two bounds with the central value for the $h \to \tau \mu$ branching fraction in Eq.~\eqref{eq:Br} leads to an upper bound
$  \mc{B}(h \to \tau e)  < 0.26$\,.
This is above the current indirect constraint coming from searches for $\tau \to e \gamma$~\cite{Aubert:2009ag} which reads $ {\mc{B}(h \to \tau e)  <0.19}$\,.
Future improvements of bounds on $\mu \to e \gamma$ and especially
$\mu-e$, and assuming $\mathcal B(h\to \tau\mu)$ stays at the percent
level, will indirectly probe $\mathcal B(h\to \tau e )$ at the
$10^{-5}$ level~\cite{Dorsner:2015mja}. Similar analysis can be carried over to the THDM III framework, where
the resulting bound on $h \to \tau e$ renders this decay invisible at the LHC
(green region in right-hand side panel in Fig.~\ref{fig:master}).

\section{Conclusions}
\label{sec:conclusions}
Motivated by the experimental indication of $h \to \tau\mu$ events by
the CMS Collaboration we have examined the implications of LFV Higgs decays 
at the percent level on several extensions of the SM. We have shown
how a tentative $\mathcal B(h\to \tau\mu)$ signal can be combined with
other Higgs measurements to yield a robust lower bound on the
effective LFV Higgs Yukawa couplings to taus and muons.
In explicit NP models, the $\tau \to \mu
\gamma$ constraint is generically more severe. In fact, an eventual
observation of $h \to \tau\mu$ at the LHC together with indirect
constraints would point in the direction of an extended SM scalar
sector, minimal example being THDM of type III. We have also examined 
models where $h \to \tau\mu$ is generated at loop level to demonstrate
difficulties with these models.
Finally, we have combined the signal of $h \to \tau\mu$ with
experimental limits on $\mu \to e \gamma$ decays and $\mu - e$
conversions in nuclei to yield robust bounds on $\mathcal B(h \to \tau
e)$. In particular, considering only the Higgs EFT effects, the two
LFV Higgs decay rates could still be comparable. On the other hand,
the THDM III cannot accommodate both branching ratios at the percent
level. It turns out that currently planned improvements in experimental searches for $\mu-e$ LFV processes will be able to probe the product $\mathcal B(h\to \tau \mu) \mathcal B(h\to \tau e)$ at the $10^{-7}$ level in generic EFT and to order $10^{-12}$ or better within the THDM III.



\printbibliography

\end{document}